\documentclass[11pt]{article}
\usepackage{moriond_noPic}
\usepackage{xspace,amssymb}
\usepackage{graphicx}

\def\tt{\ensuremath{t\overline{t}}\xspace}
\def\tbar{\ensuremath{\overline{t}}\xspace}

\begin{document}
\vspace*{4cm}
\title{TOP PHYSICS AT THE LHC}

\author{ CHRISTIAN WEISER }

\address{Institut f\"ur Experimentelle Kernphysik\\
         Universit\"at Karlsruhe\\
         }

\maketitle\abstracts{
Top quark physics will be a prominent topic in Standard Model physics
at the LHC. The enormous amount of top quarks expected to be produced
will allow to perform a wide range of precision measurements.
An overview of the planned top physics programme of the ATLAS and CMS experiments
at the LHC is given.
}

\section{Introduction}
The Large Hadron Collider (LHC) is expected to start operation in 2007.
It will accelerate proton beams and bring them to collision at
a centre of mass energy of $\sqrt{s}$=14 TeV and at luminosities
between
$(1-2) \times 10^{33} cm^{-2}s^{-1}$ (initial 'low-luminosity' phase)
and
$10^{34} cm^{-2}s^{-1}$ ('high-luminosity').
Top quark physics studies will be accomplished by the
general purpose experiments
ATLAS and CMS.

The reasons to study the physics of top quarks are numerous.
The top quark is the heaviest known fundamental particle
-- its mass being a fundamental parameter of the Standard Model --
and the only quark that does not hadronize because of its short
lifetime ($\Gamma_t>>\Lambda_{QCD}$).
The top quark mass $m_t$ is of particular interest, because it is related
through radiative corrections (the so-called $\Delta r$ parameter) to
the mass of the W and Higgs bosons, $m_W$ and $m_H$. Precise measurements of
$m_t$ and $m_W$ thus allow to constrain $m_H$.
%
%
Top quark events will also be a major source of background for many
search channels.
They can be also very useful to calibrate the detector
and reconstruction algorithms, e.g. the jet energy scales
and b-tagging performance.

At the LHC, gluon-gluon fusion is the dominant process
for \tt production, contributing about 87\% of the
production cross section, whereas quark-antiquark annihilation
contributes only about 13\%. This is contrary to the situation at
the TEVATRON.
The \tt production cross-section is about $\sigma_{\tt} \approx 800 pb$.
This implies that during the low luminosity phase of the LHC ($10^{33} cm^{-2}s^{-1}$),
about one \tt pair will be produced per second, resulting in
about $10^7$ \tt pairs per year.

Because of $|V_{tb}| \approx 1$, the top quark decays almost
exclusively into a W boson and a b quark, $BR(t \to Wb) \approx 1$.
The top quark decay is thus characterised through the decay of the
W boson.
In \tt events, about 5\% fall into the di-lepton category,
$\approx$ 30\% into the semi-leptonic one and $\approx$~45\%
into the fully hadronic one
(only electrons and muons are considered as leptons here).

\section{Measurements of the top quark mass}
The semi-leptonic channel $\tt \to b\overline{b} l\nu q\overline{q}$ is the most
promising for an accurate measurement of the top quark mass.
It has a relative large branching ratio and the isolated lepton ($e$ or $\mu$)
allows efficient triggering.

A typical analysis requires an isolated lepton, missing transverse energy,
and at least four jets within the tracker acceptance.
To suppress backgrounds and to resolve the combinatorics within the event,
one or two jets have to be tagged as b-jets.
The mass of the hadronically decaying top quark, thus the mass
of the $Wb=jjb$ system, is used for the measurement.
\begin{figure}[htb]
\begin{center}
\leavevmode
\includegraphics[width=0.34\textwidth]{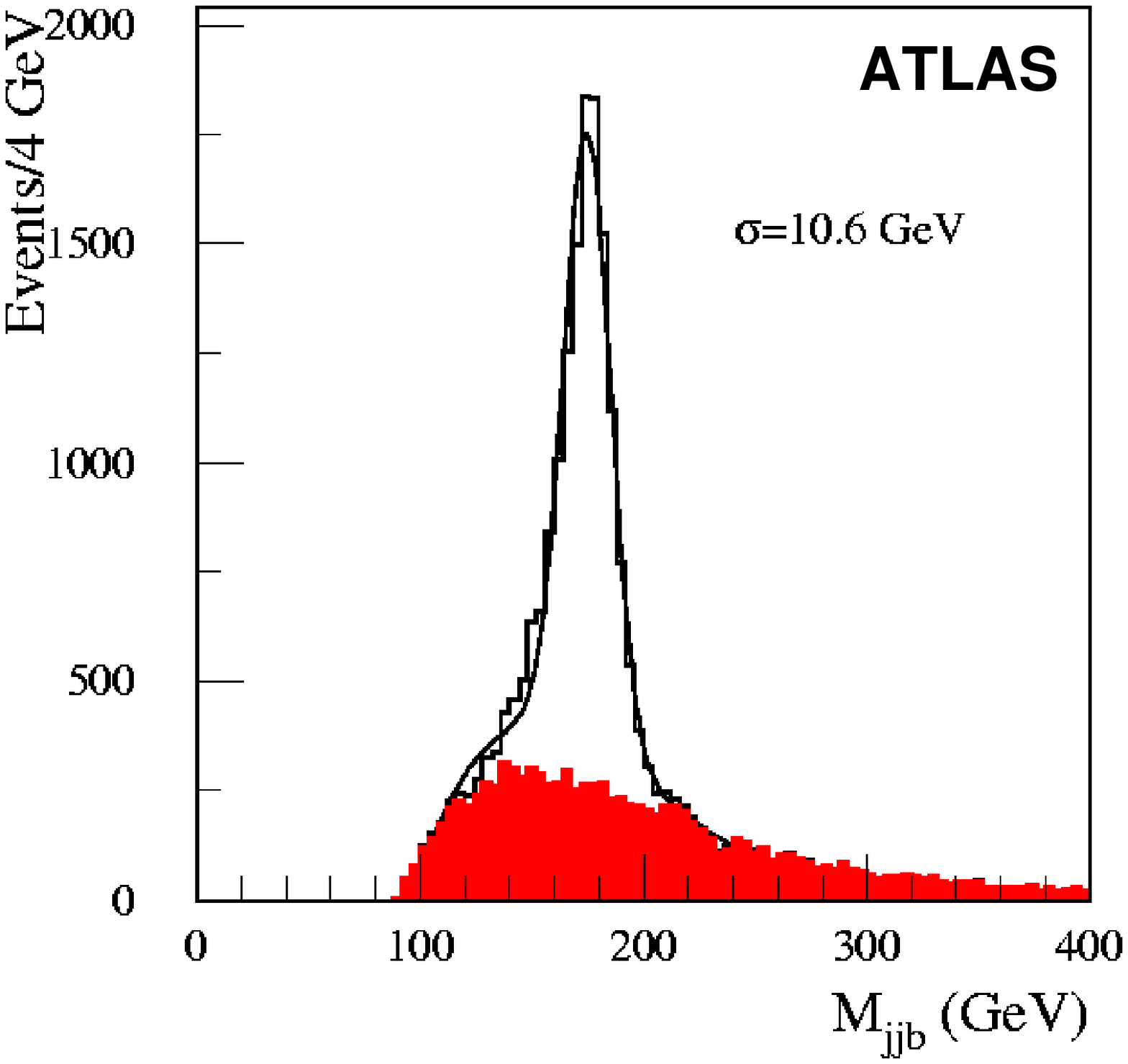}
\hspace*{0.1\textwidth}
\includegraphics[width=0.34\textwidth]{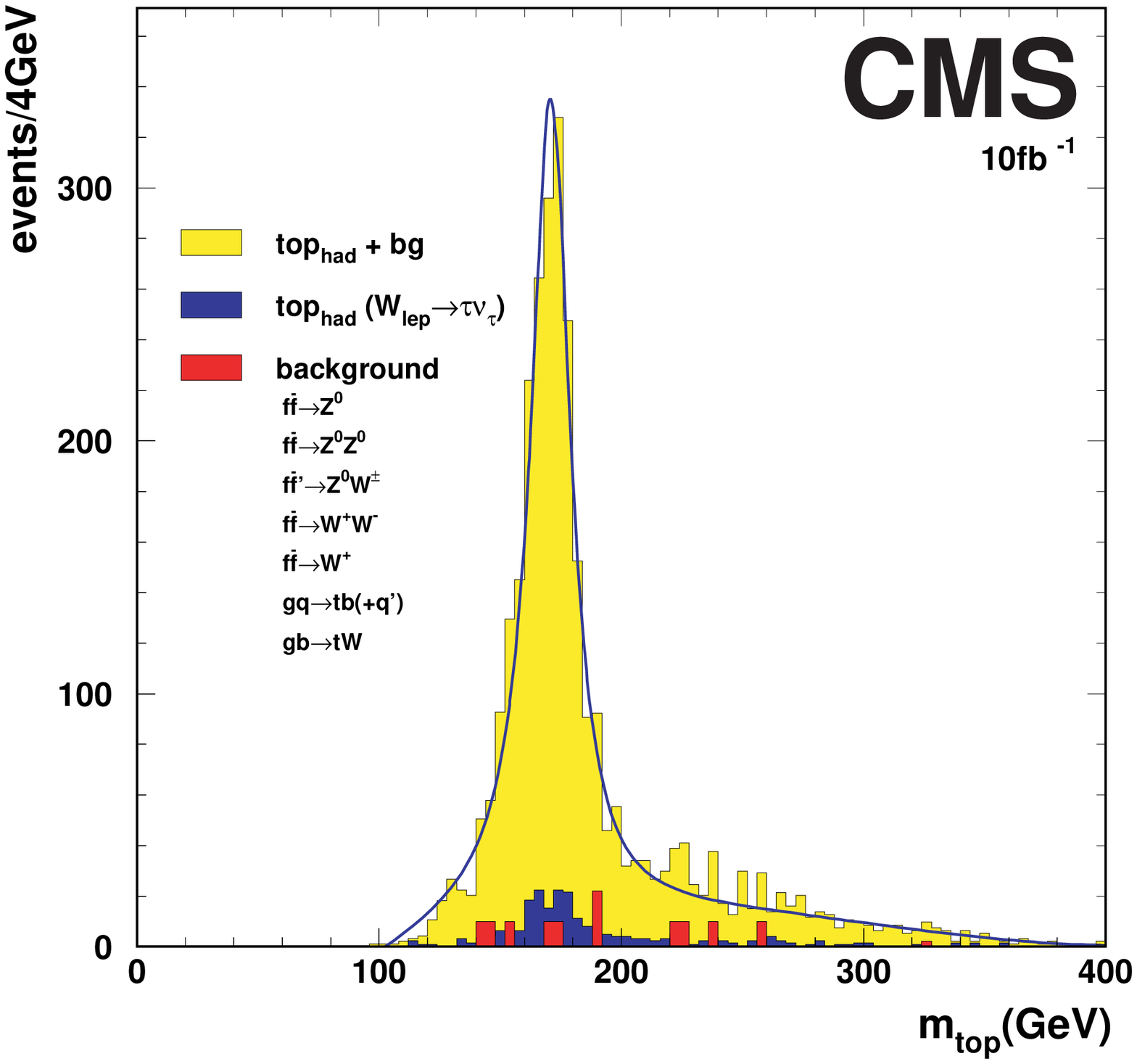}
\caption{\label{TopMassPlots}Reconstructed top mass in the semi-leptonic channel
                             for ATLAS (left) and CMS (right), for $10 fb^{-1}$ of
                             integrated luminosity. For CMS, the combinatorial background
                             within the event is included in the signal curve
                             (only non-\tt events are shown as background), whereas for
                             ATLAS it is shown in the background curve.\hspace*{1.2cm}}
\end{center}
\end{figure}

Figure \ref{TopMassPlots} shows the reconstructed top masses for
ATLAS~\cite{ATLASTopMassAll} and CMS~\cite{CMSTopMassSL}.
The remaining backgrounds (mainly W and Z boson(s) + jets) are very low.
Already after one year of running at low luminosity, the measurement uncertainty is
clearly dominated by systematics.
Of particular importance is the knowledge of the b-jet energy scale
(the sensitivity to the light jet scale is strongly reduced by using
the W boson mass as constraint), which has to be determined at the level
of a percent to reach the desired accuracy.
It seems to be feasible to reach a precision on the top mass of about 2 GeV,
maybe an ultimate precision of about 1 GeV could be reached when all effects are
well understood.

The large number of events available allows to measure the top mass
in different sub-samples using different methods showing different
sources of systematic errors.
ATLAS~\cite{ATLASTopMassAll} has studied semi-leptonic events, where the top quarks are produced
almost back-to-back with large transverse momentum, leading to very collimated
top decay products well separated in hemispheres.
The top quark mass is computed from the calorimeter towers within a large cone.
The dominant systematics using this method is then the contribution from the
underlying event and pile-up events.

The top mass can also be measured in the fully leptonic and fully hadronic channels.
The Di-Lepton channel ($e$, $\mu$) is very clean. However, the branching ratio is
quite small and because of the presence of two neutrinos in the final state
the kinematics is underconstrained.
The fully hadronic channel is experimentally very challenging.
Efficient triggering is difficult and the QCD background is enormous.
However, this channel has the biggest branching ratio ($\approx 45\%$) and
the kinematics is fully constrained.
Accuracies of about 2-3 GeV seem feasible for these channels~\cite{ATLASTopMassAll}.

ATLAS~\cite{ATLASTopMassAll} and CMS~\cite{CMSTopMassJPsi} also studied the possibility
to measure the top mass in the semi-leptonic channel in
final states with an exclusively reconstructed $J/\Psi$ from a b-hadron decay
in a b-jet, where the $J/\Psi$ carries a large fraction of the b-hadron momentum
because of its large mass.
The mass of the $J/\Psi-l$ system is strongly correlated
to the top mass.
The method is thus rather insensitive to the jet energy scale, the understanding of
the b-fragmentation becoming crucial.
Because of the tiny branching ratio of this final state, this analysis has to be carried
out during the high luminosity phase of the LHC.
Even after some 100$fb^{-1}$, the statistical error is expected to contribute significantly
to the total error ($\approx$ 1 GeV for a total error of $\approx$ 1.5 GeV).

\section{\tt production properties}
A measurement of the \tt production cross section allows to test
QCD calculations.
Differential cross sections
(e.g. $d\sigma_{\tt}/d\eta$) give access to parton distribution functions (PDFs).
Heavy particles decaying into a \tt pair could show up as resonances
in the \tt mass spectrum $d\sigma_{\tt}/dm_{\tt}$ and thus be an indication for new physics.
Furthermore, the \tt production cross section is sensitive to the top quark mass,
$\sigma_{\tt} \propto 1/m_t^2$.

Because of its short lifetime, the top quark does not form bound states and
thus there should be no dilution of the spin information in the decay.
Top quarks are not expected to be produced polarised but their spins are expected to be
correlated.
The asymmetry
${\cal A} = {
            {N(t_L\tbar_L + t_R\tbar_R) - N(t_L\tbar_R + t_R\tbar_L)}
	     \over
            {N(t_L\tbar_L + t_R\tbar_R) + N(t_L\tbar_R + t_R\tbar_L)}
	    }$
allows to distinguish between the g-g fusion and q-$\overline{\mbox{q}}$
production processes (${\cal A} \approx 0.431$ vs. ${\cal A} \approx -0.469$).
In fully leptonic events, $\cal A$ can be extracted from a fit to the double
differential distribution of the angles between the leptons in their respective top frame
and the top quark in the \tt rest frame.
A CMS study~\cite{Sonnenschein} shows that the following accuracies can be reached after 30$fb^{-1}$ of
integrated luminosity:
$\sigma({\cal{A}})_{stat.} \approx 0.035$,
$\sigma({\cal{A}})_{syst.} \approx 0.028$.

\section{Top quark properties}
The large number of \tt events will allow precise measurements of
top quark properties and tests of the V-A structure of charged current
weak interactions.

The top quark charge has not yet been experimentally confirmed to be $2/3$.
Two approaches have been studied by ATLAS~\cite{ATLASTopCharge}
to distinguish between
$t(Q=2/3) \to W^+ b$ and
$t(Q=-4/3) \to W^- b$.
The first approach investigates $\tt\gamma$ events.
Since the radiation of photons in these events is roughly
proportional to $Q_t^2$ for the production process (radiation can also
occur in the decay), the $p_t$ spectrum of the photons in $\tt\gamma$
allows to distinguish the two top quark charge values.
The second approach measures the charges of all top quark decay products.
Whereas for the leptonic W boson decay the charge is given by the charge of the
lepton ($e$ or $\mu$), a jet-by-jet estimation of the b-quark charge is not
possible.
However, quantities correlated to the charge of the b-quark allow to distinguish
between the cases on a statistical basis.
As an example, the jet charge
$Q_{jet} = { {\sum_i q_i |\vec{jet}\cdot \vec{p}_i^\kappa|} \over
             {\sum_i     |\vec{jet}\cdot \vec{p}_i^\kappa|} }$
has been studied.
Both methods should allow an unambiguous measurement of the top quark charge
after one year of data taking at low luminosity (10 $fb^{-1}$).

Because of the V-A structure of weak charged current interactions,
the W bosons in top quark decays are expected to be polarised.
The predicted population of the helicity states of the W boson are:
$h_W(-1) \approx 0.3$,
$h_W(0)  \approx 0.7$,
$h_W(+1) \approx 0$.
In semi-leptonic events, the angle between the lepton in the W boson rest frame
and the W boson in the top quark frame, $\Theta^*_l$, can be analysed.
A CMS study~\cite{Sonnenschein} gives the following accuracies for the longitudinal helicity state
after an integrated luminosity of 10 $fb^{-1}$:
$\sigma(h_W=0)_{stat.} \approx 0.023$,
$\sigma(h_W=0)_{syst.} \approx 0.022$.

\section{Single top production}
\begin{figure}[htb]
\begin{center}
\leavevmode
\includegraphics[width=0.94\textwidth]{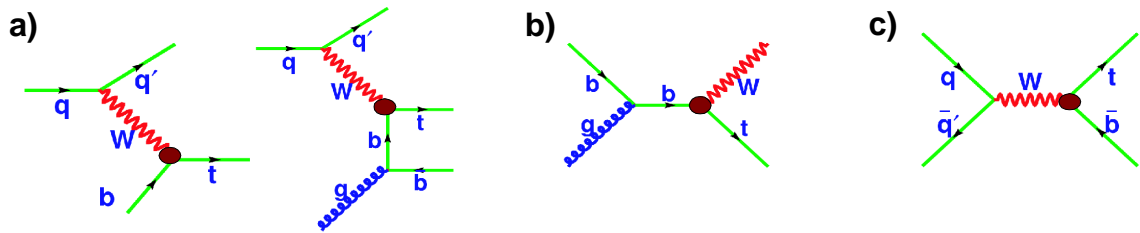}
\caption{\label{SingleTop} Single top production mechanisms at the LHC:
                           a) W-gluon fusion ($\sigma \approx 250\ pb$),
			   b) Wt production
			     ($\sigma \approx 60\ pb$)
			      and
                           c) W$^*$ or s-channel ($\sigma \approx 10\ pb$).\hspace*{6.1cm}}
\end{center}
\end{figure}

Single top quark events can be created via electro-weak processes.
The contributing diagrams together with their
expected cross-sections are shown in Figure \ref{SingleTop}.
Single top events give direct access to the absolute value of the
CKM matrix element $V_{tb}$.
Backgrounds (\tt, Wjj, Wbb) are more severe than for \tt events and
the ability to extract clear signals depends more critically on the
detector performance.
The different production processes show different final states and are
thus typically studied separately.
To reject the enormous QCD background, the leptonic decay channel
$t \to b l \nu$ ($l=e,\mu$) is studied.
An ATLAS study~\cite{ATLASPTDR} gives the following statistical experimental errors
on $|V_{tb}|$ after $30 fb^{-1}$:
$\sigma(|V_{tb}|)_{(W-g)} \approx 0.4\%$,
$\sigma(|V_{tb}|)_{(W-t)} \approx 1.4\%$,
$\sigma(|V_{tb}|)_{(W^*)} \approx 2.7\%$.
The dominating systematic uncertainties are expected to come
from theoretical uncertainties of the cross section calculations
and the luminosity measurement.

\section{Calibration with \tt events}
The possibility to isolate clean samples of \tt events allows
the calibration of reconstruction algorithms.
Using the W boson mass as a constraint allows to calibrate the jet energy
scale of light quark jets.
Jet samples enriched in b-quark jets (from the top quark decay) and light
quark jets can be isolated and used to determine the b-tagging efficiencies
on the data themselves.

\section{\tt in associated Higgs production}
For low Higgs masses ($m_H \lesssim 135$ GeV), the dominant decay
mode of the Higgs boson is $H \to b\overline{b}$.
However, this decay mode can only be exploited in channels of
associated Higgs production, $ttH$ being the most promising one.
From the experimental point of view, this channel is complementary to
the channel $H \to \gamma\gamma$.
Furthermore, this channel would allow a measurement of the
top-Higgs Yukawa coupling.
In order to find the b quark jets from the Higgs boson decay, 
the \tt system can be fully reconstructed.
ATLAS~\cite{ATLASttH} and CMS~\cite{CMSHiggs} studies indicate
that this channel might support a discovery of a light Higgs boson.

\section{Conclusions}
At the LHC, top quarks will be produced abundantly.
The goal for the top mass measurement is to reach an accuracy
of approximately 1 GeV.
The large available statistics will also allow to determine
many top quark properties both in production and decay.
Single top production will make a precise determination of
$|V_{tb}|$ possible.


\section*{References}
\begin{thebibliography}{99}
%
\bibitem{ATLASTopMassAll} I.~Borjanovi\'c et al., SN-ATLAS-2004-040 and hep-ex/0403021
%
\bibitem{CMSTopMassSL}    L.~Sonnenschein, CMS NOTE 2001/001 
%
\bibitem{CMSTopMassJPsi}  A.~Kharchilava, CMS NOTE 1999/065 
%
\bibitem{Sonnenschein} L.~Sonnenschein, PhD Thesis, RWTH Aachen, PITHA 01/04  
%
\bibitem{ATLASTopCharge} M.~Ciljak et al., ATL-PHYS-2003-035  
%
\bibitem{ATLASPTDR} ATLAS Detector and Physics Performance Technical Design Report,
                    ATLAS TDR 15 and CERN/LHCC 99-15
%
\bibitem{ATLASttH} J.~Cammin, M.~Schumacher, ATL-PHYS-2003-024;\\
                   B.~King et al., ATL-PHYS-2004-031
%
\bibitem{CMSHiggs} S.~Abdullin et al., CMS NOTE 2003/033
%
\end{thebibliography}
\end{document}